\documentclass[prl,aps,showpacs,twocolumn,unsortedaddress,longbibliography,superscriptaddress]{revtex4-2}	

\usepackage{graphics, bm, xspace}
\usepackage{graphicx}
\usepackage{psfrag}
\usepackage{amsmath}
\usepackage{amssymb}
\usepackage{epsfig}
\usepackage{subfigure}
\usepackage{grffile}
\usepackage{times}
\usepackage{color}
\usepackage{multirow}
\usepackage[bookmarks=false,linkcolor=blue,urlcolor=blue,colorlinks,citecolor=blue]{hyperref}
\usepackage{float}
\usepackage{gensymb}
\usepackage{mhchem}  

\newcommand{\beq}    {\begin{equation}}
\newcommand{\enq}    {\end{equation}}
\newcommand{\ceq}[1] {(\ref{#1})}

\newcommand{\eps}    {\epsilon}

\DeclareMathOperator {\sgn}{sgn}

\newcommand{\unity}{1\!\!1}

\newcommand{\bv}{\mathbf}
\newcommand{\peh}{T_{he}}

\begin{document}


\title{SU(4) Symmetry Breaking and Induced Superconductivity in Graphene Quantum Hall Edges}
\author{Joseph J. Cuozzo}
\author{Enrico Rossi}
\affiliation{Department of Physics, William \& Mary, Williamsburg, Virginia 23187, USA}
\date{\today}


\begin{abstract}
In graphene, the approximate SU(4) symmetry associated with the spin and valley degrees of freedom
in the quantum Hall (QH) regime is reflected in the 4-fold degeneracy of 
graphene's Landau levels (LL's). 
Interactions and the Zeeman effect break such approximate symmetry and lift the corresponding degeneracy of the LLs.
We study how the breaking of the approximate SU(4) symmetry affects the properties of 
graphene's QH edge modes located in proximity to a superconductor.
We show how the lifting of the 4-fold degeneracy qualitatively modifies the transport properties of the QH-superconductor heterojunction.
For the zero LL,
by placing the edge modes in proximity to a superconductor, it is in principle possible to realize a 1D topological
superconductor supporting Majoranas in the presence of sufficiently strong Zeeman field. 
We estimate the topological gap of such a topological superconductor and 
relate it to the properties of the QH-superconductor interface.
\end{abstract}

\maketitle

Heterojunctions formed by two-dimensional electron gases (2DEGs) in the quantum Hall (QH) regime,
placed in proximity of a superconductor (SC), are ideal to realize one-dimensional 
topological superconductors~\cite{fu2008b,Mong2014,Finocchiaro2018,SanJose2015}, and
are the only realistic systems
in which it is expected that not only Majoranas zero modes~\cite{Kitaev2001}  
but also more complex non-Abelian anyons can be realized~\cite{fu2008b,Mong2014,Clarke2013}.
In recent years advances in materials' and devices' fabrication have allowed
the realization of high quality QH-SC systems~\cite{Rickhaus2012,Amet2016,Zhao2020,Lee2017,gul2021induced,Hatefipour2022, Zhao2022}
that have shown signatures of superconducting correlations induced in the edge modes of QH states.
Such experiments 
have motivated several theoretical works~\cite{Tang2022, David2022, Kurilovich2022, Michelsen2023, Kurilovich2022_criticality, Manesco2022, Sekera2018, Beconcini2018, Hou2016, Zhang2019, Alavirad2018, Schiller2022,Galambos2022}
that addressed some of the limitations of simple models.
QH-SC systems based on graphene~\cite{Rickhaus2012,Amet2016,Zhao2020,Lee2017,gul2021induced,Zhao2022} are particularly promising for several reasons:
by encapsulating the graphene layer in hexagonal boron nitride (hBN), 
high-quality, low-disorder, devices can be realized;
they can be driven into robust QH states with smaller values of the magnetic field ($B$) 
than regular 2DEGs due 
to the fact that for 2D Dirac materials
the Landau level (LL) energies $E_n$ scale with the square root of
$B$, $E_n=\sgn(n)v_F\sqrt{2e\hbar B|n|}$ with $v_F$ the grapehene's Fermi velocity 
and $n\in\mathbb{N}$~\cite{DasSarma2011_qh,Goerbig2011}, rather than linearly with $B$,
as for 2D systems with parabolic bands.
These features have recently enabled the observation of 
superconducting correlations between the edge states of fractional QH states~\cite{gul2021induced},
the first step toward the realization of parafermions.

In graphene, due to the spin and valley degeneracy, we have an approximate 4-fold degeneracy of the fermionic states.
As a consequence, in the presence of a strong perpendicular magnetic field, graphene well approximates a $SU(4)$
quantum Hall Ferromagnet~\cite{Avoras1999,Burkov2002,Ezawa2002,Nomura2006}.
The approximate $SU(4)$ symmetry is broken by Zeeman and interactions effects~\cite{yang2006,jung2009a,abanin2013,young2012}.
The breaking of the $SU(4)$ symmetry can significantly affect the strength of the superconducting correlation
induced among QH's edge modes by proximity to a SC, and therefore modify the
conditions required for the realization of non-Abelian anyons in QH-SC systems.
In this work we study the effect that the breaking of the $SU(4)$ symmetry of graphene's Landau levels (LLs)
has on the nature and strength of electron-hole (e-h) conversion processes (Andreev reflection processes) at the interface between
the integer QH (IQH) edge and an s-wave superconductor.
For $n>0$ LLs the breaking of the 
the $SU(4)$ symmetry causes the edge modes' drift velocity ($v_d$)
to be spin and valley dependent, and we find that this 
causes the e-h conversion probability, $\peh$, to oscillate
as function of strength of the $SU(4)$ symmetry breaking terms.
For the $E_n=0$ LL the Coulomb interaction induces 
correlated phases~\cite{jung2009a,Kharitonov2012,young2012,abanin2013,Young2014,Takei2016,Wei2018,stepanov2018,liu2022,coissard2022,zhou2022a}
that lift the degeneracy between particle-like and hole-like states and so break the effective
$SU(4)$ symmetry of the LL, and we find that the interplay of interaction's effects and Zeeman splitting ($\Delta_Z$)
can strongly affect the transport properties of the $E_n=0$ QH-edge modes at a QH-SC interface. 
Our results show the effect that $SU(4)$ breaking terms have on the e-h conversion
in graphene-based QH-SC systems, and how, conversely, signatures in the transport properties of QH-SC edges can
be used to estimate the relative strength of such terms. The dependence of such transport properties 
on $\Delta_Z$ can also be used to estimate the efficiency of e-h conversion 
at QH-SC interfaces.

\begin{figure*}[ht]
  \centering
  \includegraphics[width=1.99\columnwidth]{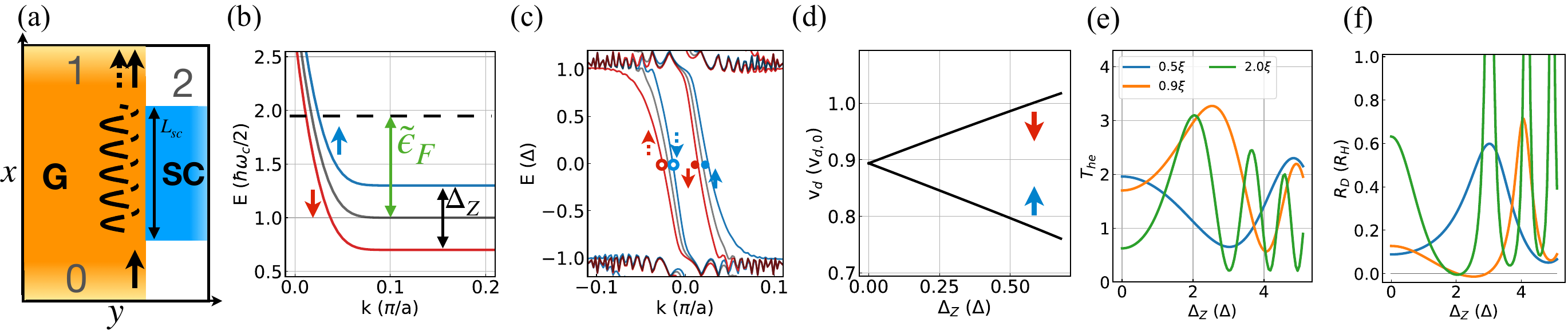}
  \caption{
   (a) Schematic of the QH-SC setup. 
   (b) Dispersion of LL with Zeeman splitting. 
   (c) BdG spectrum of a hybrid QH-SC structure in the LLL with and without (faded lines) Zeeman field. 
   (d) Drift velocity of CAESs normalized by the vacuum drift velocity versus the Zeeman field.
   $T_{he}$ (e), and $R_D$ (f), for $\tilde\epsilon_F=10~\Delta$, $v_{s} = 0.1 \Delta\xi/\hbar$, $k_F = 5 \hbar/\xi$,
   and $L_{sc}=(0.5, 0.9, 2.0)\xi$.
  }
  \label{fig.zeeman}
\end{figure*}

We consider the setup shown schematically in Fig.~\ref{fig.zeeman}(a) in which an interface of length $L_{sc}$ is present
between a SC and a graphene layer in the IQH regime. 
In this situation the IQH edge modes, along the QH-SC interface,
form a chiral Andreev edge state (CAES), a coherent superposition of e-like and h-like 
edge states~\cite{Hoppe2000, Giazotto2005, Khaymovich2010, VanOstaay2011, Lian2016}.
For graphene we assume ``armchair-like'' boundaries. Such boundaries do not
have intrinsic edge modes, like zig-zag edges~\cite{Brey2006,Wakabayashi_2010},
and therefore allow to better study the intrinsic properties of QH edge states~\cite{Brey2006}.
It is also expected that in most experimental setups 
armchair-like boundaries better approximate the devices' edges than zig-zag edges.

For $E_n>0$ LLs the $SU(4)$ symmetry breaking can be described 
by taking into account the presence of a Zeeman term of strength $\Delta_Z$ that splits the spin degeneracy
and a similar term, of strength $\Delta_{v}$, that lifts the valley degeneracy.
We consider an effective low-energy Boguliubov-de-Gennes Hamiltonian ($H_{BdG}$) for the one-dimensional (1D) edge modes
located at the interface between the QH systems and the SC. Assuming that no magnetic field is present in the SC, we have
\begin{align}
 \hat H_{BdG} = \psi^\dagger[&\hbar v_d k\tau_0\eta_0\sigma_0 -\hbar v_d k_F\tau_z\eta_0\sigma_0+(\Delta_Z/2)\tau_0\eta_0\sigma_z + \nonumber\\
                             &\hat\Delta_{K_1K_2} +\Delta\tau_x\eta_0\sigma_0]\psi
 \label{eq:HbdG}
\end{align}
where 
$
\psi=(c_{kK_1\uparrow},\allowbreak c_{kK_1\downarrow},\allowbreak c_{kK_2\uparrow},\allowbreak c_{kK_2\downarrow},\allowbreak 
      c^\dagger_{-kK_2\downarrow},\allowbreak  -c^\dagger_{-kK_2\uparrow},\allowbreak c^\dagger_{-kK_1\downarrow},\allowbreak  -c^\dagger_{-kK_1\uparrow})^T
$
with  $c^\dagger_{kK_i\sigma}$, $c_{kK_i\sigma}$
the creation, annihilation operators, respectively, 
for an electron with momentum $k$, and spin $\sigma={\uparrow,\downarrow}$, in the $K_i$ valley.
$\eps_k=\hbar v_d k$ is the edge mode's dispersion along the QH-SC interface with $v_d$ the drift velocity,
$\tau_i$, $\eta_i$, and $\sigma_i$ are $2\times2$ Pauli matrices in particle-hole, valley, and spin space, respectively, and
$k_F$ is the Fermi wavevector at the Fermi energy $E_F$. 
In Eq.~\ceq{eq:HbdG} 
$
\hat\Delta_{K_1 K_2} = \Delta_{v}/2[\cos(\theta_v)\tau_0\eta_z\sigma_0 + \sin(\theta_v)\cos(\phi)\tau_z\eta_x\sigma_0 + \sin(\theta_v)\sin(\phi)\tau_z\eta_y\sigma_0]
$,
where $\theta_v$, $\phi$ are the angles 
that parametrize the direction in valley space of the mean field lifting the valley degeneracy.
For $E_n>0$ LLs $SU(4)$ breaking terms affect the transport properties of the QH-SC edge 
by causing the edge modes' drift velocity $v_d$ to be spin and valley dependent. This is due to two mechanisms:
(i)  such terms cause the effective tunneling between the QH region and the SC to be dependent on the SU(4) flavor (see SI);
(ii) the splitting due to these terms causes edge states of different SU(4) flavors to have different Fermi wavevectors
and therefore, when the confining potential $V(y)$ is not linear, different drift velocities.
To exemplify the physics, below we consider in detail the second mechanism given that the further
inclusion of the first mechanism is straightforward and its effect is also in general quite smaller (see SI). 

To understand how $\Delta_Z$ induces a spin-dependent $v_d$ we
can consider the simple case when 
$V(y)= V_0 y^2/(l_V)^2$, for $y>0$ and $V(y)=0$ for $y<0$.
$V_0$ and $l_V$ are constants that characterize the confining potential.
Considering that $y=l_B^2 k$, with $l_B$ the magnetic length,
in the limit $dV/dy|_{\eps_k=0}\ll \hbar\omega_c/l_B$, where $\omega_c$ is the cyclotron frequency, we obtain
$v_{d\uparrow\downarrow}=v_d[\tilde\eps_F\mp\Delta_Z/2]^{1/2}$,
with $v_d=2[l_B^2/(\hbar l_V)][V_0\tilde\eps_F]^{1/2}$, 
and $\tilde\epsilon\equiv E_F-E_n$, $E_F$ being the Fermi energy, Fig.~\ref{fig.zeeman}~(b), (SI).
In the limit $\Delta_Z\ll 2\tilde\epsilon_F$ we have 
$v_{d\uparrow\downarrow}=v_d[1\pm\Delta_Z/(4\tilde\epsilon_F)]$.

We first consider
the case when $\Delta_Z\neq 0$ and $\Delta_v=0$. This situation is also directly relevant to QH-SC heterostructures
based on standard 2DEG systems.
In this case $H_{BdG}$ can be block-diagonalized with blocks 
$\hat H_{\pm} = \psi_{\pm}^\dagger H_{\pm}\psi_{\pm}$, where $H_{\pm}$ are $2\times 2$ matrices and 
$\psi_+=(c_{k\uparrow},c^\dagger_{-k\downarrow})$, $\psi_-=(c_{k\downarrow},-c^\dagger_{-k\uparrow})$.
Here we drop the valley indices since in this case the valleys are degenerate.
From the expressions of $H_\pm$ and the BdG equation $H_{\pm}\Phi_{\pm}(x) = E \Phi_{\pm}(x)$, we can calculate the transfer matrices~\cite{VanOstaay2011,Beenakker2014}
\begin{equation}
        M_{\pm}(L,~0) = e^{i\alpha} 
        \begin{pmatrix}
        t_{\pm} & \mp i a_{\pm} \\
        \mp i a_{\pm} & t^*_{\pm}
        \end{pmatrix}
        \label{eq.M}
\end{equation}
that relate the CAES's state at the end, $x=L_{sc}$, of the QH-SC interface, to the CAES's state at the beginning, $x=0$, of the interface.
In Eq.~\ceq{eq.M} $\alpha$ is a trivial phase and $a_{\pm}$ describes the mixing of electrons and holes along the interface.
Knowing $M_{\pm}$ we can obtain 
the probability for Andreev conversion $T_{he}^{(\pm)} = |a_{\pm}|^2$ of an electron with spin $+/-=\uparrow,\downarrow$ from lead 0 to lead 1 (see Fig.~\ref{fig.zeeman}~(a)):
\begin{equation}
        T_{he}^{(\pm)} =  \frac{{\Delta}^2 \sin^2\left( L_{sc} \delta k_{eh,\pm} \right)}{\left(\hbar v_s~ \delta k_{F,\pm} \right)^2}
        \label{eq.The}
\end{equation}
with
\begin{equation}
 \delta k_{he\pm} = \frac{1}{\hbar v_s}\sqrt{{\Delta}^2 + \left( \frac{\hbar v_s k_F \pm\hat{v}E-\hat{v}\Delta_Z/2}{1-\hat{v}^2}\right)^2}.
 \label{eq.khe}
\end{equation}
In Eqs.~\ceq{eq.The},~\ceq{eq.khe} 
$\hat v\equiv v_a/v_s$, with  
$v_s\equiv(v_{d\uparrow}+v_{d\downarrow})/2$, 
$v_a\equiv(v_{d\uparrow}-v_{d\downarrow})/2$.

The knowledge of $T_{he}^{(\pm)}$ allows us to obtain the resistance $R_D$
between the superconducting terminal 2 and terminal 1 in the absence of backscattering~\cite{Lee2017,Beconcini2018,gul2021induced,Hatefipour2022}. 
For filling factor $\nu$ 
we have:
\begin{equation}
 R_D  = \frac{R_H}{\nu} \frac{\left(\nu - 2 T_{he}\right)}{2 T_{he}}
 \label{eq.RD}
\end{equation}
where $R_H=h/e^2$. 
For our case, $\nu=6$ (including the $E_n = 0$ LLs) and 
$T_{he} = 2(T_{he}^{(+)}+T_{he}^{(-)}) + 2T_{he}^{(0)}$ and $T_{he}^{(0)} = T_{he}^{(+)}(E=0, v_a=0, \Delta_{v/Z}=0)$
describes the Andreev conversion of $n=0$ LL states for which asymmetries due to splittings are negligible.
Equations~\ceq{eq.The}-\ceq{eq.RD} show how the spin dependence of the edge modes velocities,
by modifying $\delta k_{he}$, affects the electron-hole conversion taking place along the QH-SC interface, and its transport properties.

To obtain a quantitative estimate of the effect of $SU(4)$-breaking terms we have also
obtained the transport properties at the QH-SC interface
using a tight-binding (TB) model implemented via the Kwant package~\cite{Groth2014}.
Details of the model can be found in section III of the SI \cite{seeSI}. 
Figure~\ref{fig.zeeman}~(c) shows the dispersion of the CAES
when $E_n>0$
both for the case when $\Delta_Z=\Delta_v=0$, and the one for which $\Delta_v=0$ but $\Delta_Z\neq 0$. 
Figure~\ref{fig.zeeman}~(d) shows the effect of $\Delta_Z$ on the renormalized, spin-dependent, drift velocity.
In the limit $\Delta_Z=0$, for the chosen parameter values, tunneling processes into the SC cause the renormalized $v_d$
to be $\sim 9/10$ of the $v_d$ at a QH-vacuum interface.
The results of Fig.~\ref{fig.zeeman}~(d) show that the scaling $v_a\propto\Delta_Z$, for $\Delta_Z\ll\tilde\epsilon_F$,
obtained assuming a quadratic edge potential agrees well with the scaling obtained from the TB-model 
calculation for $\Delta_Z$ as large as $\sim 0.5\tilde\epsilon_F$ 
Figures~\ref{fig.zeeman}~(e),~(f) show
the total e-h conversion probability, $T_{he}$, and $R_D$
as a function of $\Delta_Z$ 
for $\nu = 6$ and various $L_{sc}$, obtained assuming for $v_\uparrow(\Delta_Z)$ and $v_\downarrow(\Delta_Z)$ the scalings 
$\pm \Delta_Z/(2\tilde{\epsilon}_F)$ consistent with Fig.~\ref{fig.zeeman}~(d).
In the limit of $L_{sc}>\xi$, Fig.~\ref{fig.zeeman}~(f), we find that $T_{he}$ oscillates with $\Delta_Z$
and that the period of the oscillations decreases with $\Delta_Z$.

The results shown in Figs.~\ref{fig.zeeman}~(e),~(f) are qualitatively valid also when 
the term $\hat\Delta_{K_1K_2}$ does not break time reversal (TR) symmetry,
i.e., when $\theta_v=\pi/2$, considering that the TR operator in valley space is $\Theta_v = \mathcal{C} \eta_x$,
with $\mathcal{C}$ denoting complex-conjugation.
As a consequence, for $\theta_v=\pi/2$, $\hat\Delta_{K_1K_2}$ only affects the average value of the drift velocity and
does not induce an asymmetry between the drift velocities of the e-like and h-like states. The resulting 
transport properties at the QH-SC edge mode are obtained from Eqs.~\ceq{eq.The}-\ceq{eq.RD} by simply taking into
account the renormalization of $v_s$ due to $\hat\Delta_{K_1K_2}$.

When $\theta_v\neq\pi/2$, the term $\hat\Delta_{K_1K_2}$ breaks TR symmetry
and causes the drift velocities of the time-reversed edge modes to be different.
In this case, the effect of $\hat\Delta_{K_1K_2}$ can compound, or compensate, the effect of $\Delta_Z$.
Let 
$v_{s,K_i} \equiv (v_{d\uparrow K_i}+v_{d\downarrow K_i})/2$, 
$v_{a,K_i} \equiv (v_{d\uparrow K_i}-v_{d\downarrow K_i})/2$,
and 
$v_{s,s} \equiv (v_{s,K_1} + v_{s,K_2})/2$,
$v_{s,a} \equiv (v_{s,K_1} - v_{s,K_2})/2$,
$v_{a,s} \equiv (v_{a,K_1} + v_{a,K_2})/2$,
$v_{a,a} \equiv (v_{a,K_1} - v_{a,K_2})/2$.
Using these definitions, and setting without loss of generality $\phi=0$, 
for the case when $\theta_v=0$,
the BdG Hamiltonian, including the valley degrees of freedom, takes the form
\begin{align}
 \hat H_{BdG} &= \psi^\dagger[\hbar v_{s,s} k\tau_0\eta_0\sigma_0+ \hbar v_{s,a}k\tau_z\eta_z\sigma_0+ \hbar v_{a,s}k\tau_z\eta_0\sigma_z \nonumber \\
                             &  + \hbar v_{a,a}k\tau_0\eta_z\sigma_z 
                               + (\Delta_Z/2)\tau_0\eta_0\sigma_z + (\Delta_v/2)\tau_0\eta_z\sigma_z \nonumber \\
                             & -\hbar v_{s,s}k_F\tau_z\eta_0\sigma_0 +\Delta\tau_x\eta_0\sigma_0]\psi
 \label{eq.HbdG2}
\end{align}
\begin{figure}[h]
  \centering
  \includegraphics[width=1.0\columnwidth]{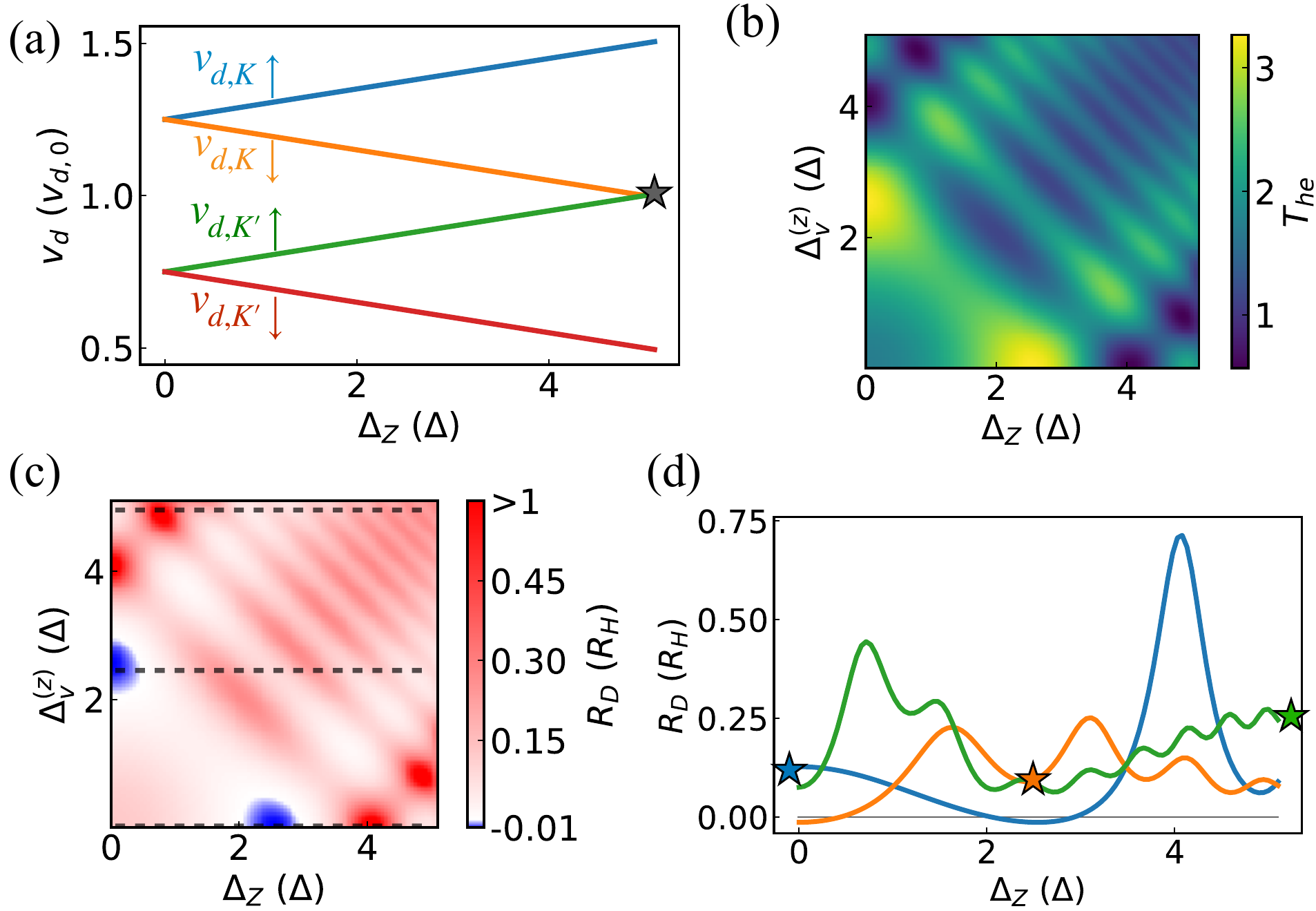}
  \caption{
   (a) Drift velocities versus $\Delta_Z$ for $\theta_v=0$, $\Delta_v = \tilde\epsilon_F /2$, and 
       $\tilde\epsilon_F=10~\Delta$. 
   (b) $T_{he}$ versus $\Delta_Z$ and $\Delta_v$ for
       $L_{sc} = 0.9~\xi$, $v_{s,s} = 0.1\Delta\xi/\hbar$, and $k_F = 5 \hbar/\xi$. 
    (c) $R_D$ corresponding to panel (b).
   (d) Line cuts of $R_D$ for fixed $\Delta_v=(0,~ 2.5,~ 5)\Delta$, blue, orange, green, traces, respectively. 
       Stars denote the values of $R_D$ when $\Delta_Z = \Delta_v$.
   }
   \label{fig.valley}
\end{figure}
Figure~\ref{fig.valley}~(a) shows the drift velocities $v_{\sigma,K_i}$ as a function
of Zeeman splitting for $\theta_v=0$ and fixed $\Delta_{v}$ obtained assuming a quadratic edge potential and $\Delta_{v,Z}\leq\tilde\epsilon_F/2$.
From the values $v_{\sigma,K_i}$, 
we can block-diagonalize Eq.~\ceq{eq.HbdG2} and calculate the momentum difference between coupled electron and hole modes. Then we use Eqs.~\ceq{eq.The} and \ceq{eq.RD} to obtain $T_{he}$ and $R_D$.
We see that as $\Delta_Z$ increases the velocity asymmetry becomes larger for a pair of CAESs while becoming 
smaller for the other pair until it vanishes when $\Delta_Z=\Delta_v$.
One could expect a maximum $\peh$ at this point, but Fig.~\ref{fig.valley}(b) shows 
that while a mirror symmetry about the $\Delta_Z = \Delta_v$ line exists, $\peh$ is not maximum along this line.
Fig.~\ref{fig.valley}~(c) shows the dependence of $R_D$ on $\Delta_Z$ and $\Delta_v$.
We find that when $\theta_v=0$, the dependence of $R_D$ on $\Delta_Z$
is different depending on the value of $\Delta_v$, as shown by the line cuts presented in Fig.~\ref{fig.valley}~(d):
the period of the oscillations of $R_D$ with respect to $\Delta_Z$ decreases as $\Delta_v$ increases.
By comparing the experimentally measured $R_D$ as a function of $\Delta_Z$, by tuning 
the in-plane component of the magnetic field, results like the ones in Fig.~\ref{fig.valley}~(d)
could allow the estimation of the 
strength of the valley splitting term breaking time reversal symmetry.
%
%
\begin{figure}[h]
  \centering
  \includegraphics[width=1.0\columnwidth]{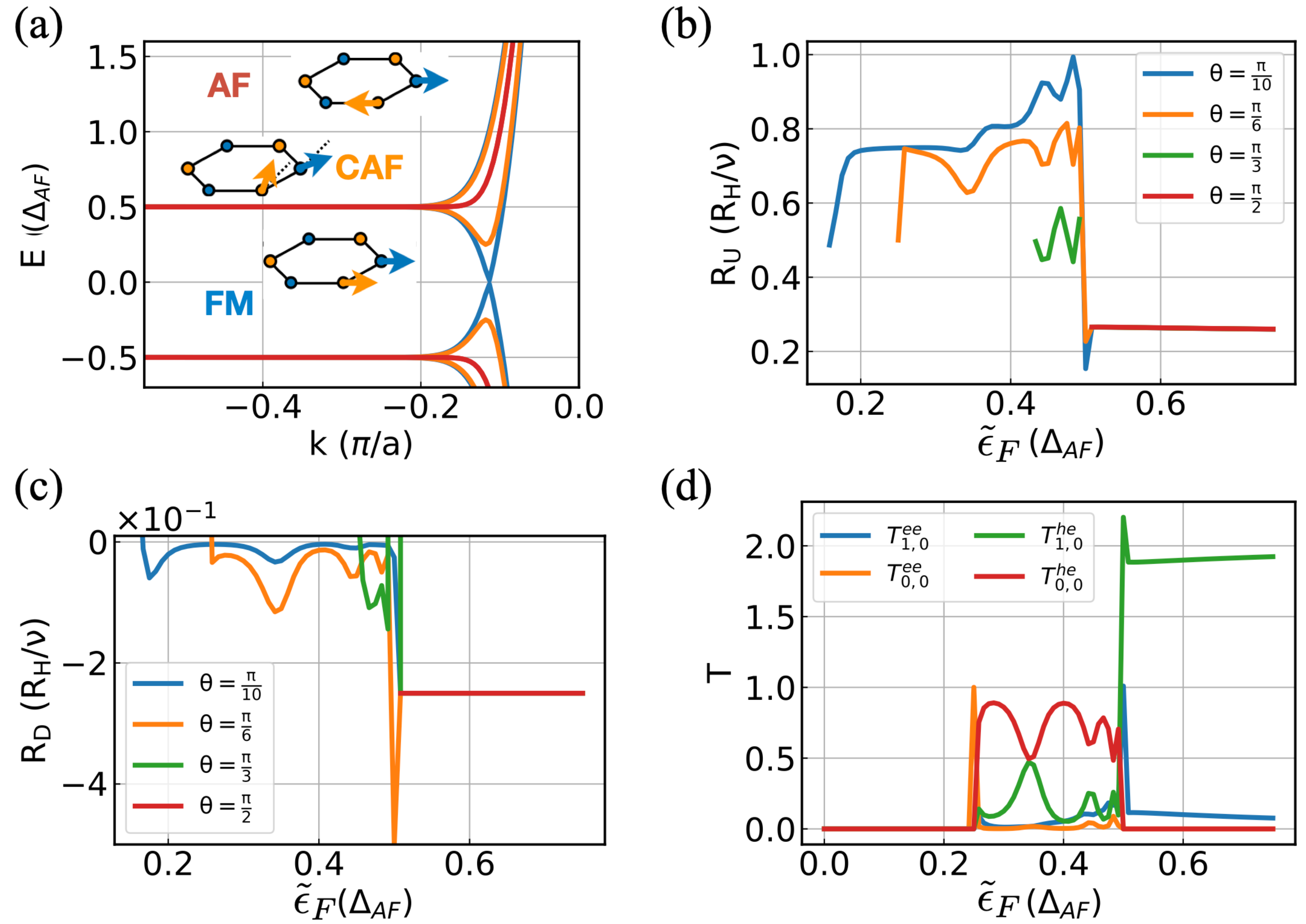}
  \caption{
   (a) Dispersion of $E_n=0$ graphene LL for the FM, CAF and AF phases with schematics illustrating the canting of spins on different sublattices for each phase.
   $R_U$ (b), and $R_D$ (c), as a function of $\tilde\epsilon_F$ in the CAF phase. 
   (d) $T^{ee}_{j,i}$ and $T^{he}_{j,i}$ versus $\tilde\epsilon_F$ for $\theta=\pi/6$. $L_{sc}=0.675\xi$.
    }
   \label{fig.n0}
\end{figure}

In the $E_n=0$ LL we have a degeneracy between particle-like and hole-like states.
However, for the $E_n=0$ LL we also have that the valley and sublattice degree of freedom are locked to each other.
Taking into account the spin degree of freedom, the full, approximate, symmetry 
for the $E_n=0$ LL is still $SU(4)$.
Besides the Zeeman effect, such approximate symmetry is broken by interaction effects
that drive the system into a correlated state. 
Theoretical~\cite{Kharitonov2012,Takei2016,das2022} and experimental results~\cite{young2012,Young2014,Wei2018,stepanov2018,zhou2022a} 
have shown that the likely correlated state is a canted antiferromagnetic (CAF) state in which the spin degree
of freedom is locked with the sub-lattice degree of freedom as shown schematically
in the inset of Fig.~\ref{fig.n0}~(a). 
Recent measurements~\cite{liu2022,coissard2022}, however, suggest that in some cases
the ground state can be an intervalley coherent phase characterized by a Kekul\'e distortion~\cite{nomura2009a}.
For $\Delta_Z$ much smaller than the interaction strength $U_{int}$, also for the $E_n=0$ LL,  
both in the CAF and Kekul\'e phase, 
the effect of $\Delta_Z$ 
on the transport properties at the QH-SC interface is described  by Eqs.~\ceq{eq.M}-\ceq{eq.RD}.
When $\Delta_Z$ is comparable to, or larger than, $U_{int}$ 
its effect on the transport properties of the QH-SC interface can be significantly  
different from the one described by Eqs.~\ceq{eq.M}-\ceq{eq.RD}, but qualitatively
the same in the Kekul\'e phase and CAF phase.
In the remainder we focus on the CAF phase.

For $\Delta_Z\ll U_{int}$ the CAF phase corresponds to an
antiferromagnetic (AF) state, whereas for $\Delta_Z\gg U_{int}$
the CAF phase describes a ferromagnetic 
(FM) state~\cite{Young2014}, see inset Fig.~\ref{fig.n0}~(a).
To describe the CAF phase, 
to the tight-binding Hamiltonian for graphene 
(section IV SI) we add the term
$
 H_{AF}  = (\Delta_{AF}/2) [
             \sum_{i}  (\psi_{A_i}^{\dagger} \tau_z \sigma_x  \psi_{A_i}
                       -\psi_{B_i}^{\dagger} \tau_z \sigma_x  \psi_{B_i})]
$
where $\psi_{S_i}^{\dagger}$, $\psi_{S_i}$, are the creation, annihilation, operators
for an electron at site $S_i=(A_i,B_i)$ with $A_i$, $B_i$ the sites of sublattices $A$, $B$, respectively,
and $\Delta_{AF}$ the strength of the mean-field describing the AF phase. 
The term $H_{AF}$, without Zeeman splitting, induces a bulk and edge gap at the charge neutrality point, as seen in Fig.~\ref{fig.n0}~(a).

To describe the evolution from the AF phase to the FM phase,
we set $\Delta_{AF}=\Delta_0\sin\theta$ and $\Delta_{Z}=\Delta_0\cos\theta$,
where $\Delta_0\equiv [\Delta_{AF}^2+\Delta_Z^2]^{1/2}$ is the total magnitude of the bulk gap,
and $2\theta$ is the angle between the spin projections on sublattice A and B.
Figure~\ref{fig.n0}~(a) shows the evolution of the LL close to the neutrality point as $\theta$ is varied:
for $\theta=0$ we recover the FM phase, and for $\theta=\pi/2$ the AF phase. 
When $\Delta_Z\neq 0$ the lowest energy particle-like and hole-like states approach close to the edge
causing the gap between edges states ($\Delta_{edge}$) to be smaller than the gap between bulk states,
Fig.~\ref{fig.n0}~(a). 
For $\Delta_Z\neq 0$, close to the edge, the spin polarization becomes momentum dependent so that forward and backward moving modes have opposite
spin polarizations (see SI). As a consequence, when $\Delta_Z\sim\Delta_{AF}$, 
for the QH edge in proximity of the SC we can have strong Andreev retroreflection.

Figures~\ref{fig.n0}~(b),~(c) show the calculated resistance $R_U$,
between terminal 2 and 0, and
$R_D$, respectively, as a function of $\tilde\epsilon_F$ for the $E_n=0$ LL in the CAF phase 
for different values of $\theta$. 
Given that for $\Delta_Z\gtrsim \Delta_{AF}$ we can have counter-propagating modes,
the equations for $R_D$, Eq.~\ceq{eq.RD}, and $R_U$ have to be generalized to
take into account backscattering processes, see Fig.~\ref{fig.n0}~(d) and SI.
As $\theta$ increases $R_U$ decreases. This fact 
could be used to extract the effect of interactions on the dispersion of the $E_n=0$ LL's edge modes.

\begin{figure}[h]
  \centering
  \includegraphics[width=1.0\columnwidth]{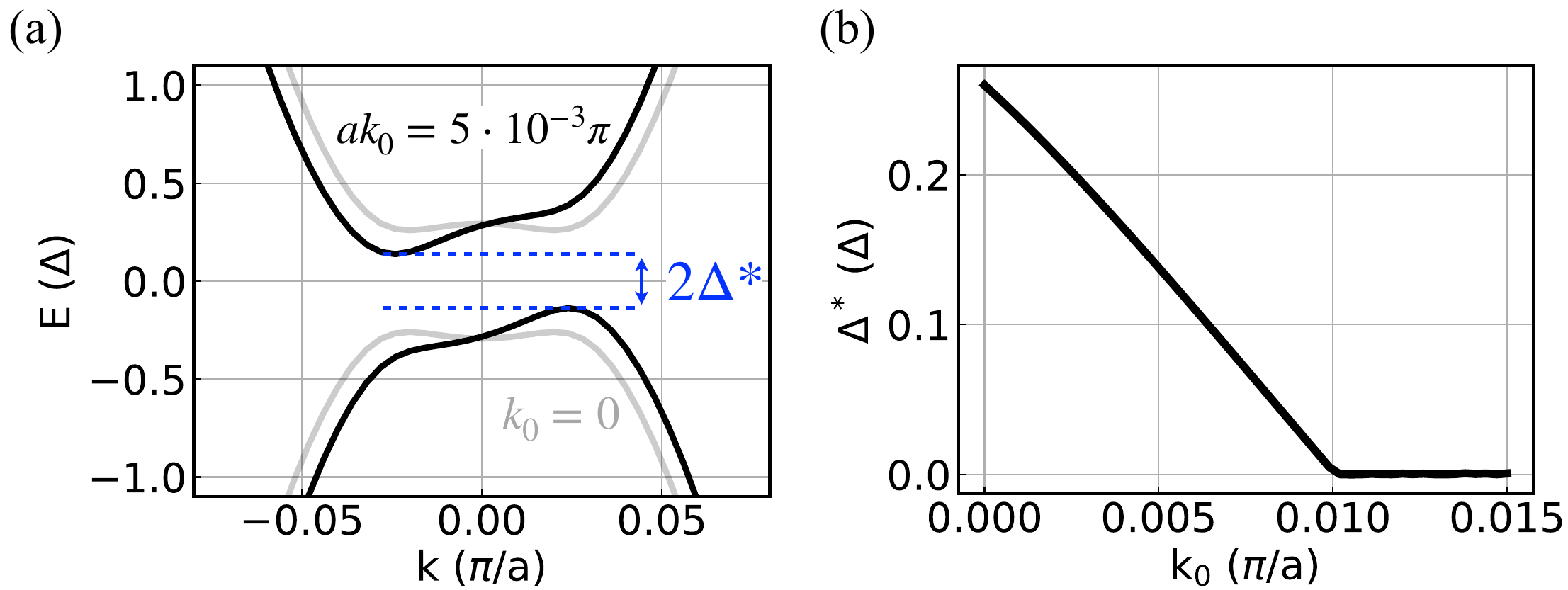}
  \caption{    
   (a)  Spectrum for effective 1D model with finite superconducting pairing $\Delta$ and small offset $k_0$.
   (b)  Topological gap $\Delta^*$ as a function of $k_0$ for a fit corresponding to $\theta = \pi/4$. 
    }
   \label{fig.1d}
\end{figure}

For $\Delta_Z\gtrsim\Delta_{AF}$, as we approach the FM phase, 
the $n=0$ LL has counter-propagating 
edge modes with non-trivial spin structure that 
appear to be ideal for the realization of
a 1D topological superconducting state supporting Majoranas at its 
ends~\cite{SanJose2015}.
However, using realistic parameters' values, and the full graphene-SC TB model,
we find that no Majoranas are present.
To understand the issue
we map the CAF edge states 
to the effective 1D model 
$
H = t \hat k^2-\epsilon_F + \alpha \hat k \sigma_y + J_Z \sigma_z
$
with $\hat k \equiv a k$, $a=2.46\AA$ being graphene's lattice constant, 
and extract 
$t$ and $\alpha$ from fitting the LL's edges dispersion obtained from the tight-binding model~\cite{seeSI}.
For $\theta=\pi/4$, using the TB model parameters presented in section IV of the SI, we find
$t= 63 \Delta$, $\alpha = 7 \Delta$.
Given that for the CAF regime considered we only have one helical band we
can set $\epsilon_F=0$ and set $J_Z$ equal to the chemical potential of the single helical band.
In the remainder we set $J_Z=4\Delta$.  
The key difference between the ideal 1D model and the edge at the QH-SC interface is that 
for the latter the minimum of the bands when $\alpha=0$, in general, is not located at 
a time-reversal invariant momentum ($k=0$), that for a QH-SC system corresponds to the edge between QH region and SC. 
To take this into account in the effective 1D model we introduce a momentum offset $k\to k+k_0$.
As $k_0$ increases the band-gap $\Delta^*$ induced by the superconducting paring becomes more indirect and is reduced, as can be seen in
Fig.~\ref{fig.1d}. 
We see that $\Delta^*$ vanishes when $k_0\approx 10^{-2}\pi/a$. 
Considering that 
for $B=2$~T $l_B\approx 18$~nm, we have that $\Delta^*$ will be vanishing when
the distance between QH edge modes and the SC edge is larger than $10^{-2}(\pi/a)l_B^2\approx 2.3 l_B$. 
The fact that for realistic parameters' values $k_0> 10^{-2}\pi/a$ (see SI), explains why in our TB calculations
no Majorana are observed, and points to an aspect that must be taken into account in experiments.

In summary, we have studied how the breaking of the approximate SU(4) symmetry of graphene's Landau levels
affects the Andreev conversion processes of QH edges states located in proximity of a superconductor. 
We have found that in general the probability of an electron to be converted into a hole
while traveling along the QH-SC interface, due to Andreev processes, can strongly oscillate
as a function of the strength of the terms breaking the SU(4) symmetry,
inducing oscillations of directly measurable transport properties
that could be used to extract the efficiency of the electron-hole conversion,
and the magnitude of the SU(4) breaking terms. 
Considering in detail the case when the $E_n=0$ state is in the canted-antiferromagnetic phase,
we have obtained how the canting angle of such phase
qualitatively affects the electron-hole conversion probability,
and therefore could be estimated by measuring the transport properties at the QH-SC interface.
In the limit of large Zeeman splitting
the edge modes of the $E_n=0$ LL 
have all the properties to allow the realization, when proximitized by a SC,
of 1D topological superconducting states with Majoranas. We have shown how the topological gap 
of such state could be much lower than na\"ively expected due to the nature of the edge modes' dispersion at the QH-SC interface.
Our results are directly relevant to the ongoing effort to induce superconducting pairing correlations
in graphene QH edges state with the ultimate goal to realize non-abelian anyons.

\paragraph{Acknowledgements} The work was supported by ARO (Grant No. W911NF-18-1-0290). 
E. R. also acknowledges partial support from DOE, grant No. DE-SC0022245, 
and thanks the Aspen Center for Physics, which is supported by NSF Grant No. PHY-1607611, 
where part of this work was performed, and Urlich Z\"ulicke and Yafis Barlas for helpful discussions.


\bibliography{qhsc}


\clearpage
\newpage

\renewcommand{\thefigure}{S\arabic{figure}}
\setcounter{figure}{0}    

\renewcommand{\theequation}{S\arabic{equation}}
\setcounter{equation}{0}  

\newpage
\onecolumngrid

\begin{center}
\large{\bf SUPPLEMENTAL INFORMATION}
\end{center}


\section{I. Flavor-dependent Drift Velocity due to non-linear confining potential}
Let's consider the generic Hamiltonian describing a free-electron gas in the two-dimensional (2D) $(x,y)$ plane
\begin{align}
    H = \hat\epsilon(k_x, k_y) + V(y)\unity - E_F\unity,
    \label{eq.si.H}
\end{align}
where $\hat\epsilon(k_x,k_y)$ is a matrix, in orbital space,
describing the dispersion of the 2D electron system, ${\bf k}=(k_x,k_y)$ is the 2D wavevector,
$V(y)$ is the confining potential defining the edge of the sample, and $E_F$ is the Fermi energy. 
Let's assume that translational symmetry is preserved along the $x$ direction.
In this case to include the effect of a magnetic field $B$ perpendicular 
to the 2D electron system it is convenient to use the gauge
${\bf A}=(-By,0,0)$. By replacing in \ceq{eq.si.H} ${\bf k}=(k_x,k_y)$ with ${\bf k}-(e/c){\bf A}$,
we obtain the energy spectrum 
\begin{align}
    E(k) = E_n + V(kl_B^2)-E_F,
    \label{eq.si.ek}
\end{align}
where $\{E_n\}$ are the energies of the Landau levels, $k\equiv k_x$, and $l_B=[\hbar c/(eB)]^{1/2}$ is the magnetic length.
Equation~\ceq{eq.si.ek} is valid for the common situation when $\partial_y V(y) \ll (E_{n+1}-E_n)/{l_B}$.
$\{E_n\}$ are momentum independent and so
the drift velocity of an edge state in the $n^{th}$ Landau level is determined by the confining potential:
\begin{align}
    v_d = \frac{1}{\hbar}\frac{d E_n}{dk} = \left.\frac{1}{\hbar}\frac{d V}{dk}\right|_{k_F}=\left.\frac{l_B^2}{\hbar}\frac{d V}{dy}\right|_{y_F}
\end{align}
where $k_F$ is the Fermi wavevector ($E(k_F)=0$), and $y_F=k_Fl_B^2$.
A very natural and physical approximation for the confining potential $V(y)$ is:
\begin{align}
    V(y) = \begin{cases}
    (V_0/l_V^2) \left( y - L_0 \right)^2, & y \ge L_0 \\
    (V_0/l_V^2) \left( y + L_0 \right)^2, & y \le -L_0 \\
    0, & -L_0<y<L_0.
    \end{cases}
\end{align}
where $V_0$ (with units of energy) $l_V$ (with units of length) are constants that parametrize the dependence of $V$ on the position, 
and $2L_0$ is the width of the QH system.
For the states at the Fermi energy on the $y>0$ side of the sample we have:
\begin{align}
    y_F = L_0 + l_V\sqrt{\frac{\tilde\epsilon_F}{V_0}},
\end{align}
where 
$\tilde\epsilon_F\equiv E_F-E_n$, so that
\begin{align}
    v_d = \left.\frac{l_B^2}{\hbar}\frac{d V}{dy}\right|_{y_F} = \frac{2 l_B^2}{\hbar l_V} \sqrt{V_0\tilde\epsilon_F}.
\end{align}

In the presence of a Zeeman term we have the spin-degeneracy of the energy levels $E(k)$ is lifted and we have:
$E(k) = E_n + V(kl_B^2)-E_F \pm \Delta_Z / 2$
and therefore two different Fermi wavevectors, one for the spin-up state ($k_{F+}$) and one of the spin-down state  ($k_{F-}$), 
and correspondingly, two different values of $y_F$:
\begin{align}
    y_{F,\pm} = L_0 + l_V \sqrt{\frac{\tilde\epsilon_F \mp \Delta_Z/2}{V_0}}.
\end{align}
The fact that $y_{F,+}\neq y_{F,-}$ implies that when $V(y)$ is not linear the spin-up and spin-down state
will have different drift velocities: 
\begin{align}
    v_{d\uparrow\downarrow} & = \frac{2 l_B^2}{\hbar l_V}  \sqrt{V_0 \left( \tilde\epsilon_F \mp \Delta_Z/2 \right)}
    \approx v_d \left(1 \mp \frac{\Delta_Z}{4 (\tilde\epsilon_F)} \right) = v_d \mp v_{a}, \label{eq.si.vdz}
\end{align}
where the approximate expression is valid when $\Delta_Z \ll 2\tilde\epsilon_F$, 
and the $-$, $+$ signs apply to the  spin $\uparrow$ and $\downarrow$ states, respectively. 
Thus, the Zeeman effect in combination with a nonlinear confining potential results in a nonzero
difference $v_a$ between $v_{d\uparrow}$, and $v_{d\downarrow}$.
For the case when a term is present that lifts graphene's valley degeneracy, the same equation~\ceq{eq.si.vdz}
is obtained if such a term also breaks time-reversal symmetry. In this case $\Delta_Z$ should be replaced
by $\Delta_v$, the strength of the valley-splitting term.


\section{II. Flavor-dependent Drift Velocity due to spin-dependent QH-SC tunneling strength}
Consider a chiral edge state in the lowest Landau level propagating in the x-direction. The low-energy BdG description of a spinful Landau level is
\begin{align}
    H_{qh} = \frac{1}{2} \sum_k \Psi_k^{\dagger} \left(v_d k \unity - E_F \tau_z   \sigma_0\right) \Psi_k = \frac{1}{2} \sum_k \Psi_k^{\dagger} ~h_{qh}(\vec{k})~ \Psi_k,
\end{align}
where $v_d$ is the drift velocity, $\Psi_k^{\dagger} = (c_{k\uparrow}^{\dagger},~c_{k\downarrow}^{\dagger},~c_{-k\uparrow},~c_{-k\downarrow})$ is the BdG spinor, and $\tau_i$ and $\sigma_i$ are Pauli matrices in Nambu and spin space, respectively. We take the units where $\hbar = 1$. Suppose we couple this system to an s-wave spin-singlet superconductor described by
\begin{align}
    H_{sc} = \frac{1}{2} \sum_{\bv{k}} \Phi^{\dagger}_{\bv{k}} \left[\xi_{\bv{k}} \tau_z   \sigma_0 - \Delta \tau_y   \sigma_y \right] \Phi_{\bv{k}} = \frac{1}{2} \sum_{\bv{k}} \Phi^{\dagger}_{\bv{k}} ~h_{sc}(\bv{k})~ \Phi_{\bv{k}},
\end{align}
    where $\Delta$ is the superconducting gap (assumed to be real for simplicity), $\xi_{\bv{k}} = \frac{k^2}{2m} - E_{F,s}$, and $\Phi_{\bv{k}}^{\dagger} = (d_{\bv{k}\uparrow}^{\dagger},~d_{\bv{k}\downarrow}^{\dagger},~d_{-\bv{k}\uparrow},~d_{-\bv{k}\downarrow})$. We will describe the coupling between the two systems within the tunneling Hamiltonian description with
\begin{align}
    H_{T} = \frac{1}{2} \sum_{\bv{k}} \Phi_{\bv{k}}^{\dagger} \left(t_0 \tau_z   \sigma_0 \right) \Psi_{k} + h.c. = \frac{1}{2} \sum_{\bv{k}} \Phi_{\bv{k}}^{\dagger}~h_{T}~ \Psi_{k} + h.c.,
\end{align}
where $t_0$ is the tunneling amplitude associated with electron scattering from the quantum Hall sample to the superconductor and vice versa. 
The bare Green's function for the superconductor at $T=0$ is
\begin{align}
    G_{sc}(\bv{k},~\omega) = & \left(\omega - h_{sc}(\bv{k}) \right)^{-1} \\
    = & \frac{1}{\omega^2 - \xi_{\bv{k}}^2 - \Delta^2} \left( \omega \unity + \xi_{\bv{k}} \tau_z \sigma_0 - \Delta \tau_y  \sigma_y \right).
\end{align}
The self-energy is given by
\begin{align}
    \Sigma(\bv{k},~\omega) &= \int d\bv{q}~ h_{T}(\bv{q}) G_{sc}(\bv{k}+\bv{q},\omega) h_T(-\bv{q}) \nonumber \\
                            &\approx - \lambda \frac{\omega \unity + \Delta \tau_y   \sigma_y}{\sqrt{\Delta^2 - \omega^2}}.
    \label{eq.sigma}
\end{align}
where $\lambda = -\pi t_0^2 N_{int}(0)$ and $N_{int}(0)$ is the interface DOS at the Fermi energy.
In the limit $\omega \ll \Delta$ from the equation for the poles of the dressed Green's function
\begin{align}
    \mathrm{Det}\left( h_{qh}(\bv{k}) + \Sigma(\bv{k},~\omega) - \omega \right) = 0.
\end{align}
we can obtain the effective Hamiltonian~\cite{stanescu2011,Alavirad2018}
\begin{align}
    h_{eff}(k) = \frac{v_d}{1 + \lambda/\Delta} k \unity - \frac{E_F}{1 + \lambda/\Delta} \tau_z \sigma_0 - \frac{\Delta}{1 + \Delta/\lambda} \tau_y   \sigma_y,
\end{align}

\begin{center}
\begin{figure}[h!!!]
    \centering
    \includegraphics[width=0.5\textwidth]{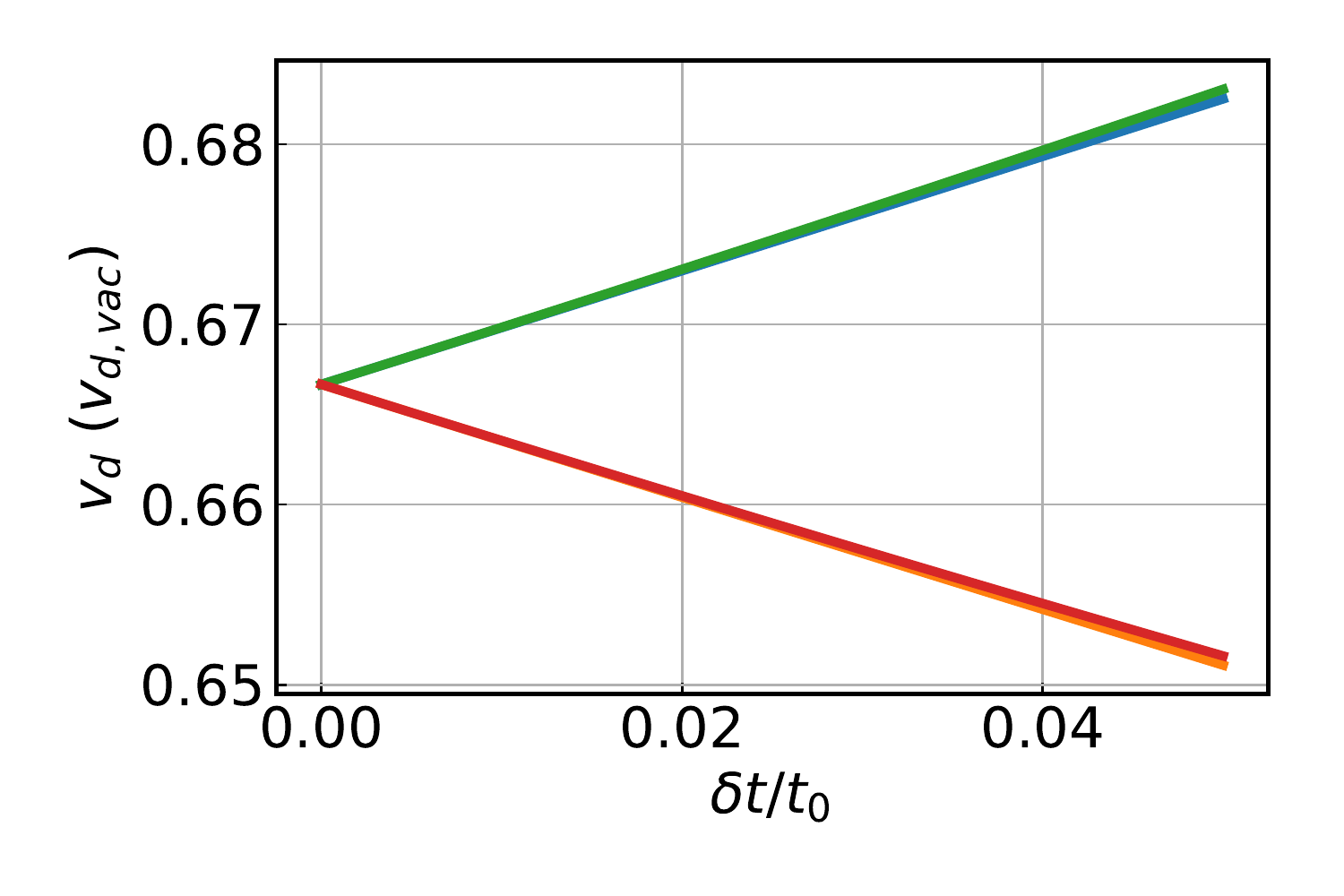}
    \caption{Drift velocity of edge states as a function of $\delta t$ for $E_F = \Delta/2$, $\lambda = \Delta / 2$, $\Delta_Z = 0$, and $v_d = 5\cdot 10^2 a \Delta$.
    }
    \label{fig:dt_dep}
\end{figure}
\end{center}

Now we will consider a perturbation to the tunneling Hamiltonian. Consider the Hamiltonian for a vacuum edge state with Zeeman splitting $\Delta_Z$:
\begin{align}
    H^{\prime}_{qh} = \frac{1}{2} \sum_k \Psi_k^{\dagger} \left(v_d k \unity - E_F \tau_z   \sigma_0 + \frac{\Delta_Z}{2}\tau_z   \sigma_z \right) \Psi_k =
                      \frac{1}{2} \sum_k \Psi_k^{\dagger} h'_{qh}(k)\Psi_k
\end{align}
Besides the Zeeman effect lifting the degeneracy of the Landau levels, the splitting also spatially separates spin-polarized edge states with opposite spin \cite{Chklovskii1992}. This separation occurs in the direction perpendicular to the boundary (y-direction in this case). Thus, in a QH/SC heterostructure in the LLL, one edge state moves closer to the interface and the other moves further away. To account for this spatial shift, we consider the modified tunneling Hamiltonian,
\begin{align} 
    H^{\prime}_{T} = \frac{1}{2} \sum_{\vec{k}} \Phi_{\bv{k}}^{\dagger} \left(t_0 \tau_z   \sigma_0 + \delta t(\Delta_Z)~ \tau_z   \sigma_z \right) \Psi_{k} + h.c.
\end{align}
where $\delta t = \delta t(\Delta_Z)$ is an odd function in $\Delta_Z$. Now we proceed as before, where we solve for the interface self-energy with the modified tunneling Hamiltonian. Doing this, we find
\begin{align}
    \Sigma^{\prime}(\bv{k},~\omega) & = \int d\bv{q}~ h^{\prime}_{T}(\bv{q}) G_{sc}(\bv{k}+\bv{q},\omega) h^{\prime}_T(-\bv{q}) \\
    & \approx \Sigma(\bv{k},\omega) - 2 \lambda \frac{\delta t}{t_0} \left(\frac{\omega \tau_0   \sigma_z}{\sqrt{\Delta^2 - \omega^2}} \right) + \mathcal{O}\left((\delta t/t_0)^2\right).
\end{align}
Then the energy eigenvalues describing the Andreev edge states are determined by the BdG equation
\begin{align}
    \mathrm{Det}\left( h^{\prime}_{qh}(k) + \Sigma^{\prime}(\omega) - \omega \right) = 0.
\end{align}
Assuming the low-energy case $\omega \ll \Delta$, we have
\begin{align}
    \mathrm{Det}\left[ v_d k \unity - E_F \tau_z \sigma_0 - \lambda \tau_y \sigma_y 
    + \frac{\Delta_Z}{2}\tau_z\sigma_z - \tau_0\left(\sigma_0  
    + \frac{\lambda}{\Delta} \left(1 + 2\frac{\delta t}{t_0}~\sigma_z\right) \right) \omega \right] \approx 0.
    \label{eq:Heff_dt}
\end{align}
The matrix structure of the equation does not allow one to easily extract an effective Hamiltonian description. The equations for the eigenenergies are also cumbersome, so we will simply point out the primary effect of interest to us. The electron-like dispersion will have the form
\begin{align}
    E_{\pm}(k) = a_0 + a_1 k \pm \sqrt{f(k)},
\end{align}
where $a_0,~a_1\in \mathbb{R}$ are constants and $f(k)$ is a second-order polynomial in $k$. Then the velocities of these modes are
\begin{align}
    \frac{dE_{\pm}}{dk} = a_1 \pm \frac{1}{2\sqrt{f(k)}} \frac{df}{dk},
    \label{eq.t-asymm}
\end{align}
Figure~\ref{fig:dt_dep} shows the drift velocities calculated directly 
using Eqs.~(\ref{eq:Heff_dt})-\ceq{eq.t-asymm}. 
For the physically relevant regime when $\delta t /t_0 \ll 1$ we see that $v_a$ is quite small and grows linearly with $\delta t /t_0$.

%


\section{III. Tight Binding Model for 2DEG-superconductor junction}
To estimate the properties of the CAES dispersion, for $\nu=2$ we use the following tight-binding Bogoliubov-de Gennes Hamiltonian:
\begin{align}
   H_{BdG}  =& \sum_{i} \psi^{\dagger}_i \left( 4t - \mu_i \right)\tau_z\sigma_0 \psi_{i} 
   + \sum_{\left< ij \right>} \psi^{\dagger}_i \left( -te^{i\phi_{i,j}} \frac{\tau_0 + \tau_z}{2} + t e^{-i\phi_{i,j}}\frac{\tau_0 - \tau_z}{2} \right)\sigma_0      
   \psi_{j} \nonumber \\
   &+ \frac{\Delta_Z}{2}\sum_{i}\psi_{i}^{\dagger}\tau_0\sigma_z \psi_{i}
    +\sum_{i}  \psi_{i}^{\dagger} \left(-\Delta_i \tau_x \right) \psi_{i}
   \label{eq:HTB}
\end{align}
where $\psi_i = (c_{i,\uparrow},~c_{i,\downarrow},~c^{\dagger}_{i,\downarrow},~-c^{\dagger}_{i,\uparrow})^T$,
$c^{\dagger}_{i,\sigma}$ ($c_{i,\sigma}$) is the creation (annihilation) operator for an electron at site $i$ with spin $\sigma$,
$\tau_i$ are $2\times 2$ Pauli matrices in particle-hole space, $\mu_i$ and $\Delta_i$ are the chemical potential and superconducting gap at site $i$, respectively,
and $\phi_{i,j}$ is the Peierls phase introduced to take into account the presence of the magnetic field in the 2DEG.
We assume $\Delta_i=\tilde\Delta$ in the SC and $\Delta_i=0$ in the 2DEG.
In the 2DEG, $\mu_i$ is set to a value, $\mu= \hbar \omega_c$ (between the first and second Landau levels). 
We take the hopping parameter $t=1.323$ eV and lattice spacing $a=2.0$ nm to model a quadratic dispersion with effective mass $m^*=0.1 m_e$. We set $\tilde\Delta=1$~meV and consider lead widths $L_x^{(n)} = 200$ nm and $L_x^{(n)} = 600$ nm for the normal and superconductor leads, respectively.
In the SC $\mu_i=\mu_s$, and in general $\mu_s\neq\mu$.
The magnetic field is in the direction, $z$, perpendicular to the $xy$ plane to which the 2DEG is confined: ${\bf B}=B{\hat z}$.
Using the Landau gauge ${\bf A}=Bx \hat{e}_y$ (assuming translational invariance of the leads in the y-direction),
the Peierls phase is given by the expression:
\begin{equation}
        \phi_{i,j} = -\frac{2 \pi B}{\phi_0} \frac{(y_i+y_j)(x_j - x_i)}{2},
        \label{eq.si.peirls}
\end{equation}
where $(x_i,y_i)$ are the coordinates of the $i^{\mathrm{th}}$ site, $\phi_0 = h/e$ is the quantum Hall magnetic flux quantum.


\section{IV. Tight Binding Model for graphene-superconductor junction}

We consider a three-terminal graphene device with two QH leads (lead 0 and 1) and a single SC lead. The tight-binding Hamiltonian is given by
\begin{equation}
    H = H_0 + H_{sc} + H_Z
\end{equation}
In graphene each unit cells is formed by two carbon atoms, $A$, and $B$. Atoms $A$ and $B$ form two triangular lattices.
Let $A_i$, and $B_i$ denote the positions of atoms $A$ and $B$, and $S_i=A_i, B_i$. With this notation
we can write: 
\begin{align}
    H_0 & = \sum_{\left< S_i S_j \right>} \psi^{\dagger}_{S_i} \left( -te^{i\phi_{S_i,S_j}} \frac{\tau_0 + \tau_z}{2} + 
    t e^{-i\phi_{S_i,S_j}}\frac{\tau_0 - \tau_z}{2} \right) \sigma_0 \psi_{S_j} \nonumber \\
    &- \mu \sum_{S_i} \psi_{S_i}^{\dagger} \tau_z \sigma_0 \psi_{S_i}, \\
    H_{sc} & = \sum_{S_i}  \psi_{S_i}^{\dagger} \left(-\Delta \tau_y \sigma_y \right) \psi_{S_i}, \\
    H_Z & = \sum_{S_i}  \psi_{S_i}^{\dagger} \left(\frac{\Delta_Z}{2} \tau_z \sigma_z \right) \psi_{S_i},
    \label{eq.si.tbgr}
\end{align}
where $\psi_{S_i} = (c_{S_i,\uparrow},~c_{S_i,\downarrow},~c^{\dagger}_{S_i,\uparrow},~c^{\dagger}_{S_i,\downarrow})$, 
$c^{\dagger}_{S_i,\sigma}$ ($c^{\dagger}_{S_i,\sigma}$) is the creation (annihilation) operator
for an electron with spin $\sigma=\uparrow,\downarrow$ at site $S_i$,
$t = 2.8$~eV~\cite{DasSarma2011_qh},
$\tau_i$ and $\sigma_i$ are Pauli matrices in Nambu and spin space, respectively, 
$\mu$ is the chemical potential with respect to the charge neutrality point,
and $\phi_{S_i,S_j}$ is the Peierls phase.
Using the same gauge as discussed in the previous section we have
\begin{equation}
    \phi_{S_i,S_j} = -\frac{2 \pi B}{\phi_0} \frac{(S_{i_y}+S_{j_y})(S_{j_x} - S_{i_x})}{2},
\end{equation}
where $(S_{i_x},S_{j_y})$ are the coordinates of site $S_i$.
The magnetic field used in our simulations is $B=200$ T, which is artificially large to compensate for a small scattering region giving us a magnetic flux comparable to experiment ($\Phi_{tot}/\Phi_0 \approx 345$) and to fix the ratio $l_B/\xi_{sc} = 0.175$
and $L_{sc}/\xi = 0.675$. 

\begin{center}
\begin{figure}[h!!!]
    \centering
    \includegraphics[width=0.3\textwidth]{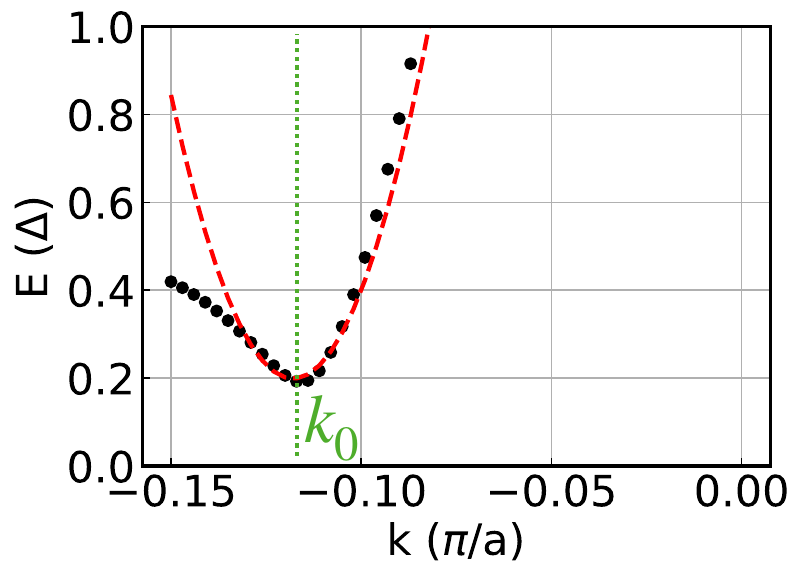}
    \caption{(Left) Parabolic fit (red dashed line) of $\nu = 0$ CAF states (black dots) for $\theta = \pi/4$ generated from the GS tight binding model using parameters used in Fig.~3 of the main text. 
    }
    \label{fig:CAF_fit}
\end{figure}
\end{center}

We use this tight binding model as the basis for the fit of $\nu = 0$ states in the CAF phase. In Fig.~\ref{fig:CAF_fit} we show the fit for $\theta = \pi /4$ used to generate Fig.~4 in the main text and confirm that the nanowire model for the $k_0 = 0$ case hosts Majorana zero modes at the ends.

%

%
%
\section{V. Calculation of transport properties}
After implementing the tight-binding model via the Python package Kwant \cite{Groth2014},
we obtain the local and non-local conductances transmission probabilities 
$T_{j,0}^{pe} = \sum_{n}^{\mathrm{N_{mode}}} t_{j,0}^{pe}(n)$ 
where $t_{j,0}^{pe}(n)$ 
is the probability of an incident electron from the $n^{th}$ band of lead $0$ to be 
scattered to lead $j$ as an electron or hole ($p=e,h$). 
The non-local downstream conductance is 
\begin{equation}
 \mathrm{G_D} = \sum_{n}^{N_{modes}}\left(T_{1,0}^{ee}(n) - T_{1,0}^{he}(n)\right)
\end{equation}
and the non-local Andreev conductance is 
\beq
 \mathrm{G_{AR}} = \sum_{n}^{N_{modes}} \left(T_{0,0}^{he}(n) - T_{1,0}^{he}(n)\right).
\enq
The upstream and downstream resistances are calculated using a Landauer-B\"{u}ttiker approach and found to be~\cite{Beconcini2018}:
\begin{align}
    R_U & = \frac{R_H}{\nu} \left(\frac{2T_{1,0}^{he} + T_{0,1}^{he} + T_{0,1}^{ee}}{2D} \right) \\
    R_D & = \frac{R_H}{\nu} \left(\frac{T_{1,0}^{ee}  - T_{1,0}^{he}}{2D} \right)
\end{align}
where
\begin{align}
    D = T_{0,1}^{he} T_{1,0}^{ee} + T_{0,1}^{ee} T_{1,0}^{he} + T_{0,0}^{ee} \left( T_{1,1}^{he} + T_{0,1}^{he} + T_{0,1}^{ee} \right) + T_{1,1}^{he} \left( T_{0,0}^{he} + T_{1,0}^{he} + T_{1,0}^{ee} \right)
\end{align}


\section{VI. On the Peierls Substitution in QH-SC Junction Simulations}
\begin{figure}[h]
    \centering
    \includegraphics[width=0.8\textwidth]{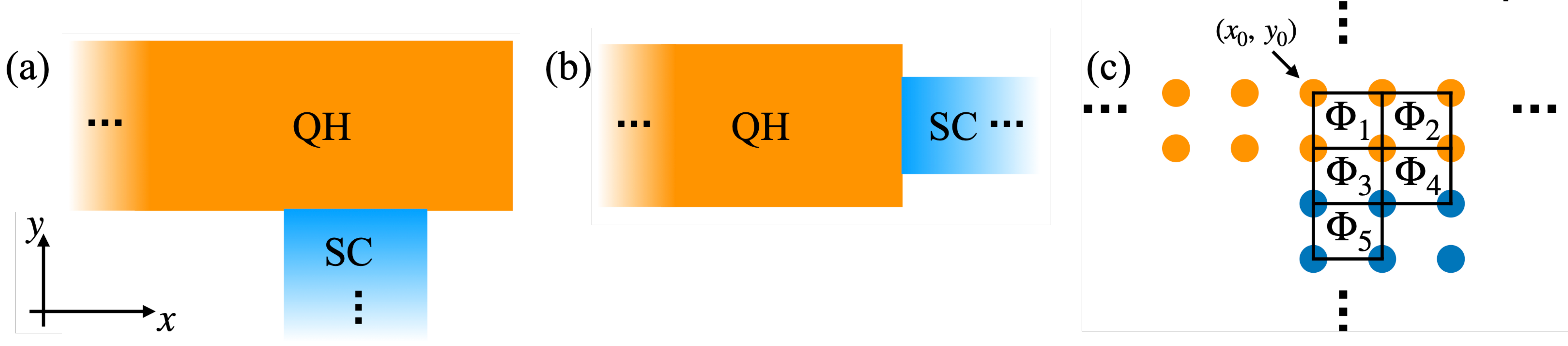}
    \caption{(a,b) Two distinct geometries of a NS junction where the two-terminal conductance is expected to be the same for both geometries in the ideal situation. (c) Schematic of the tight binding lattice at the NS interface.}
    \label{fig:NS_geometries}
\end{figure}

Let's consider the two geometries shown in Fig.~\ref{fig:NS_geometries}(a-b). 
The magnetic field can be accounted for on a lattice by using the Peierls substitution on the hopping $t$ in tight binding simulations:
\begin{align}
    \sum_{\left< ij \right>} \psi^{\dagger}_i \left( -t \right) \psi_i \rightarrow \sum_{\left< ij \right>} \psi^{\dagger}_i \left( -te^{i\phi_{i,j}} \right)\psi_i,
\end{align}
where $\phi_{i,j}$ is the Peierls phase, Eq.~\ceq{eq.si.peirls}.
The Peierls phase, integrated around a plaquette, is the magnetic flux threading the plaquette modulo $\phi_0$.
We will assume that the S region is in the Meissner phase and take $\vec{A} = 0$. First consider geometry (a) and the flux threading the plaquettes along the NS interface shown in Fig.~\ref{fig:NS_geometries}(c). Using $\Phi = \oint d\vec{\ell} \cdot \vec{A}$,
\begin{align}
    \Phi_1 & = \Phi_2 = a^2 B \\
    \Phi_3 & = \Phi_4 = a (y_0 - a)B \\
    \Phi_5 & = 0.
\end{align}
Notice that $\Phi_3,~\Phi_4$ depend on the coordinate system (i.e. $y_0$). Since $\Phi$ is an observable quantity, this cannot be the case. The choice of $y_0$ is made to smoothly and monotonically take $\Phi \rightarrow 0$ across the NS interface. Then we must choose $y_0 \in \{a,~2a \}$. But the reason for the choice of $y_0$ is even more basic than this. From classic electrostatics, the boundary conditions for the magnetic field imply the vector potential must be continuous across any boundary. Then, since the NS boundary lies between $y \in \{y_0 - a,~ y_0 - 2a \}$, we must have $y_0 \in \{a,~2a \}$ to make $\vec{A}$ continuous across the NS interface.

Turning to geometry (b), we can perform a similar analysis and show that the continuity of $\vec{A}$ along the NS interface is generally \textit{violated}. Hence, we may assume $\vec{A} = 0$ at all S sites \textit{only if} (i) the NS interface is perpendicular to the translationally-invariant normal leads and (ii) the coordinate system is chosen such that $\vec{A}$ in the normal region goes to zero at the NS interface. 

\end{document}